\documentclass[11pt,english]{article}
\usepackage[T1]{fontenc}
\usepackage[latin9]{inputenc}
\usepackage[letterpaper]{geometry}
\usepackage{amsmath}
\usepackage{amssymb}

\makeatletter
\def\eea{\end{eqnarray}}\def\be{\begin{equation}}\def\ee{\end{equation}}\def\bef{\begin{figure}}\def\enf{\end{figure}}\def\part{\partial}\newcommand{\bea}{\begin{eqnarray}}\newcommand{\ena}{\end{eqnarray}}\newcommand{\beano}{\begin{eqnarray*}}\newcommand{\enano}{\end{eqnarray*}}\newcommand{\bee}{\begin{enumerate}}\newcommand{\ene}{\end{enumerate}}\newcommand{\bei}{\begin{itemize}}\newcommand{\eni}{\end{itemize}}\newcommand{\bs}{\begin{slide}}\newcommand{\es}{\end{slide}}\def\rg{\rangle }

\makeatother

\usepackage{babel}

\makeatother

\usepackage{babel}

\begin{document}

\title{Coherent state quantization of paragrassmann algebras}

\author{M. El Baz%
\thanks{Laboratoire de Physique Théorique, LPT-URAC 13, Faculté des Sciences,
Université Mohamed V, Av.Ibn Battouta, B.P 1014 Agdal Rabat, Morocco,
E-mail: elbaz@fsr.ac.ma%
}, R. Fresneda%
\thanks{Instituto de Física, Universidade de São Paulo, Caixa Postal 66318-CEP,
05315-970 São Paulo, S.P., Brazil E-mail: fresneda@gmail.com%
}, J.P. Gazeau%
\thanks{Laboratoire APC, Université Paris Diderot \emph{Paris 7} 10, rue A.
Domon et L. Duquet 75205 Paris Cedex 13, France, E-mail: gazeau@apc.univ-paris7.fr%
} and Y. Hassouni%
\thanks{Laboratoire de Physique Théorique, LPT-URAC 13, Faculté des Sciences,
Université Mohamed V, Av.Ibn Battouta, B.P 1014 Agdal Rabat, Morocco,
E-mail: y-hassou@fsr.ac.ma%
}}
\maketitle
\begin{abstract}
By using a coherent state quantization of paragrassmann variables,
operators are constructed in finite Hilbert spaces. We thus obtain
in a straightforward way a matrix representation of the paragrassmann
algebra. This algebra of finite matrices realizes a deformed Weyl-Heisenberg
algebra. The study of mean values in coherent states of some of these
operators leads to interesting conclusions.

\bigskip{}


\smallskip{}

\end{abstract}

\section{Introduction}

In this paper, we present the quantization of a paragrassmann algebra
through appropriate coherent states \cite{csquant}. paragrassmann
algebras were introduced mainly within the context of parastatistics,
fractional statistics and more. For a quite comprehensive review of
the mathematics involved in such a structure, we refer for instance
to \cite{filisakur,isaev}. Generalizations of Grassmann algebras
have been very popular during the nineties (see e.g. \cite{ruspi}-\cite{majrod}
and references therein) from different points of view. Those generalizations
were mainly motivated by searches of exact solutions or at least integrability
properties found in 2$d$ conformal field theories, anyonic models
and topological field theories which led to unusual statistics (parastatistics,
fractional, braid statistics, ...). There were also attempts to generalize
supersymmetry to parasupersymmetry.

In Section \ref{csq}, we explain the principle and the basics of
coherent state construction on an arbitrary measure space and the
issued quantization, which we call coherent state (CS) quantization
throughout this paper. CS quantization extends to a wider set of situations
the Klauder-Berezin-Toeplitz (called also anti-Wick) quantization
\cite{ber75,ab3,lieb73}. In Section \ref{prg} we recall the main
features of the one-variable paragrassmann algebra equipped with an
integral calculus in the sense of Berezin-Majid \cite{ber65,majrod}.
In Section \ref{prgcs}, we build the corresponding coherent states
as they have been introduced in \cite{daoudkibler,rabat1,rabat2,Daoud1998}.
In Section \ref{qtferm}, we proceed with the quantization of the
one-variable paragrassmann algebra through these coherent states and
describe in detail the matrix algebra obtained from this procedure.
In Section \ref{lowsymb}, we examine the \emph{dequantization} that
is implemented via the mean values of matrices in coherent states
({}``lower symbols'') and which leads to the original paragrassmann
algebra. In Section \ref{dvarprg}, we generalize the procedure to
$d$-variable paragrassmann algebra. 
Eventually, we give in the conclusion some insights on the use of
such a formalism in finite quantum mechanics and possibly in quantum
information.

\section{Coherent state quantization}

\label{csq}

Let $X=\{x\,\mid\, x\in X\}$ be a set equipped with a measure $\mu(dx)$
and $L^{2}(X,\mu)$ the Hilbert space of square integrable functions
$f(x)$ on $X$ : \begin{equation}
\Vert f\Vert^{2}=\int_{X}\vert f(x)\vert^{2}\,\mu(dx)<\infty\,,\quad\langle f_{1}|f_{2}\rangle=\int_{X}\overline{f_{1}(x)}f_{2}(x)\,\mu(dx)\,.\label{norminprod}\end{equation}
 Let us select, among elements of $L^{2}(X,\mu)$, an orthonormal
set $\mathcal{S}_{N}=\{\phi_{n}(x)\}_{n=0}^{N-1}$, $N$ being finite
or infinite, which spans, by definition, a separable Hilbert subspace
$\mathcal{K}$ in $L^{2}(X,\mu)$. We demand this set to obey the
following crucial condition \begin{equation}
0<\mathcal{N}(x)\equiv\sum_{n}\vert\phi_{n}(x)\vert^{2}<\infty\ \mbox{almost everywhere}.\label{factor}\end{equation}
 This condition is obviously trivial for finite $N$. Let $\mathcal{H}$
be a separable Hilbert space with orthonormal basis $\{|e_{n}\rg\,,\, n=0,1,\dotsc,N-1\}$
in one-to-one correspondence with the elements of the set $\mathcal{S}_{N}$.
Then consider the family of states $\{|x\rangle\,,\, x\in X\}$ \underline{in}
$\mathcal{H}$ obtained through the following linear superpositions:
\begin{equation}
\left|x\right\rangle \equiv\frac{1}{\sqrt{\mathcal{N}(x)}}\sum_{n}\overline{\phi_{n}(x)}\left|e_{n}\right\rangle .\end{equation}
 This defines an injective map (which should be continuous w.r.t some
minimal topology affected to $X$ for which the latter is locally
compact): \[
X\ni x\mapsto\left|x\right\rangle \in\mathcal{H},\]
 By construction, these \textit{coherent} states obey 
\begin{itemize}
\item \textbf{Normalization} \begin{equation}
\langle\, x\,|x\rangle=1,\label{norma}\end{equation}

\item \textbf{Resolution of the unity in $\mathcal{H}$} \begin{equation}
\int_{X}|x\rangle\langle x|\,\mathcal{N}(x)\,\mu(dx)=\mathbb{I}_{\mathcal{H}},\label{resunit}\end{equation}

\end{itemize}
A \textit{classical} observable is a function $f(x)$ on $X$ having
specific properties, depending on supplementary structure allocated
to set $X$. Its coherent state quantization consists in associating
to $f(x)$ the operator \begin{equation}
A_{f}:=\int_{X}f(x)|x\rangle\langle x|\,\mathcal{N}(x)\,\mu(dx).\label{oper}\end{equation}
Operator $A_{f}$ is symmetric if $f(x)$ is real-valued, and is bounded
if $f(x)$ is bounded. The original $f(x)$ is an {}``upper symbol''
in the sense of Lieb \cite{lieb73} or a contravariant symbol in the
sense of Berezin \cite{ber75} for the usually non-unique operator
$A_{f}$. It will be called a \emph{classical} observable if its {}``lower
symbol'' in the sense of Lieb \cite{lieb73} or its covariant symbol
in the sense of Berezin \cite{ber75}, $\check{A}_{f}(\alpha)\overset{def}{=}\left\langle x\right|A_{f}\left|x\right\rangle $,
has mild functional properties to be made precise (e.g. is a smooth
function) according to further topological properties given to the
original set $X$ (e.g. is a symplectic manifold).

\textit{Such a quantization of the set $X$ is in one-to-one correspondence
with the choice of the frame of coherent states encoded by the resolution
of the unity (\ref{resunit}). To a certain extent, a quantization
scheme consists in adopting a certain point of view in dealing with
$X$ (compare with Fourier or wavelet analysis in signal processing).
Here, the validity of a precise frame choice is asserted by comparing
spectral characteristics of quantum observables $A_{f}$ with data
provided by some specific protocol in the observation of $X$. }

\section{The essential of paragrassmann algebra}

\label{prg} In this paper, we consider as an observation set $X$
the paragrassmann algebra $\Sigma_{k}$ \cite{filisakur,ruspi,majrod,daoudkibler}.
We recall that a Grassmann (or exterior) algebra of a given vector
space $V$ over a field is the algebra generated by the exterior (or
wedge) product for which all elements are nilpotent, $\theta^{2}=0$.
paragrassmann algebras are generalizations for which, given an integer
$k>2$, all elements obey $\theta^{k^{\prime}}=0$, where $k^{\prime}=k$
for odd $k$ and $k^{\prime}=k/2$ for even $k$. For a given $k$,
we define $\Sigma_{k}$ as the linear span of $\left\{ 1,\dotsc,\theta^{n},\dotsc,\theta^{k^{\prime}-1}\right\} $
and of their respective \emph{conjugates} $\bar{\theta}^{n}$ : here
$\theta$ is a paragrassmann variable satisfying $\theta^{k^{\prime}}=0$.

Variables $\theta$ and $\bar{\theta}$ do not commute: \begin{align}
 & \theta\,\bar{\theta}=q_{k}\bar{\theta}\,\theta\,,\label{ncparagr}\end{align}
 where $q_{k}=q=e^{\frac{2\pi i}{k}}$ for odd $k$, and $q_{k}=q^{2}$
for even $k$. The motivation backing the above definition is to choose
the deformation parameter in the paragrassman algebra as a root of
the unity of the same order as the nilpotency order: $q^{k}=1$ for
odd $k$, and $\left(q^{2}\right)^{k/2}=1$ for even $k$.

Thus, the number $k$ fixes the order of the root $q$ and the order
of nilpotency of the paragrassmann algebra. The distinction between
even and odd values of $k$ regarding the nilpotency order is necessary
to enforce that the Fock representations of the quantized paragrassmann
algebras have dimension $k^{\prime}$, and therefore match the representation
theory of the $q$-oscillator algebra when $q$ is a $k$th root of
unity (see, e.g., \cite{Chaichian}).

A measure on $X$ is defined as \begin{equation}
\mu(d\theta d\bar{\theta})=d\theta\, w(\theta,\bar{\theta})\,\, d\bar{\theta}\,.\label{dkmeas13}\end{equation}
 The integral over $d\theta$ and $d\bar{\theta}$ should be understood
in the sense of a Berezin-Majid-Rodríguez-Plaza integral \cite{majrod}:
\begin{equation}
\int d\theta\,\theta^{n}=0=\int d\bar{\theta}\,\bar{\theta}^{n}\,,\quad\mbox{for}\ n=0,1,\dotsc,k^{\prime}-2,\label{bermajint13}\end{equation}
 with \begin{equation}
\int d\theta\,\theta^{k^{\prime}-1}=1=\int d\bar{\theta}\,\bar{\theta}^{k^{\prime}-1}\,.\label{bermajintA13}\end{equation}
 The {}``weight'' $w(\theta,\bar{\theta})$ is given by the $q$-deformed
polynomial \begin{equation}
w(\theta,\bar{\theta})=\sum_{n=0}^{k^{\prime}-1}\left[n\right]_{q}!\,\theta^{k^{\prime}-1-n}\bar{\theta}^{k^{\prime}-1-n}\,.\label{qdefpol13}\end{equation}
 We adopt in this paper the {}``symmetric'' definition%
\footnote{In many places, the $q$-deformed numbers are defined in asymmetric
way as : \begin{equation}
\left[x\right]_{q}:=\frac{1-q^{x}}{1-q}\,,\label{qdefpolA130}\end{equation}
 and so are complex. 
} for $q$-deformed numbers which makes them real: \begin{equation}
\left[x\right]_{q}:=\frac{q^{x}-q^{-x}}{q-q^{-1}}=\frac{\sin\frac{2\pi x}{k}}{\sin\frac{2\pi}{k}}\,.\label{qdefpolA13}\end{equation}
 The $q$-factorial function is defined by \begin{equation}
\quad\left[n\right]_{q}!=\left[1\right]_{q}\left[2\right]_{q}\cdots\left[n\right]_{q},\,\,\text{with}\quad\left[0\right]_{q}!=1\,.\label{qfact}\end{equation}
 The measure (\ref{dkmeas13}) allows one to define a complex pseudo-Hilbertian
structure on the vector space that is the $k^{2}$-dimensional linear
span on $\mathbb{C}$ of all monomials of the type $\theta^{m}\,\bar{\theta}^{\bar{m}}$
for $m,\,\bar{m}=0,1,\dotsc,k^{\prime}-1$. A generic element of this
algebra, element of the pseudo-Hilbert space $L_{w}^{2}\equiv L^{2}(X,\mu(d\theta d\bar{\theta}))$
will be denoted: \begin{equation}
v(\theta,\bar{\theta})=\sum_{m,\bar{m}}v_{m\bar{m}}\theta^{m}\,\bar{\theta}^{\bar{m}}\,,\quad v_{m\bar{m}}\in\mathbb{C}\,.\label{elprghil}\end{equation}
 Consistently, the inner product in $L_{w}^{2}$ is defined by \begin{equation}
(v,v')=\iint d\theta\,:\,\overline{v(\theta,\bar{\theta})}\, v'(\theta,\bar{\theta})\, w(\theta,\bar{\theta})\,:\, d\bar{\theta}\,,\label{inprod}\end{equation}
 where the symbol {}``$:\,\cdot\,:$'' means an antinormal ordering,
i.e., all nonbarred $\theta$ stand on the left. In general, the monomials
$\theta^{m}\,\bar{\theta}^{\bar{m}}$ are not orthogonal. Only pairs
$\left(\theta^{m}\,\bar{\theta}^{\bar{m}},\theta^{m'}\,\bar{\theta}^{\bar{m'}}\right)$
for which $m-\bar{m}\neq m'-\bar{m'}$, are orthogonal. Note that
we have the same occurrence in the Hilbert space $L^{2}\left(\mathbb{C},e^{-\vert z\vert^{2}}\, d^{2}z/\pi\right)$
with monomials $z^{m}\,\bar{z}^{\bar{m}}$. In the latter case, an
orthonormal basis is provided by complex Hermite polynomials \cite{ghanmi,intissar}.
An adequate $q$-deformation of these complex Hermite polynomials
could provide an orthonormal basis for $L_{w}^{2}\equiv L^{2}(X,\mu(d\theta d\bar{\theta}))$.

Thanks to the inner product in \eqref{inprod}, we can define a pseudo-norm:
\begin{equation}
\vert\vert v(\theta,\bar{\theta})\vert\vert^{2}=\iint d\theta\,:\,\overline{v(\theta,\bar{\theta})}\, v(\theta,\bar{\theta})\, w(\theta,\bar{\theta})\,:\, d\bar{\theta}\,.\label{qnorm}\end{equation}
 As expected, this quantity is not positively defined on the entire
pseudo-Hilbert space $L_{w}^{2}$; on the other hand, it is positively
defined on the linear span of powers of $\theta$ and $\bar{\theta}$
separately; \textit{i}.e. on functions of the form \begin{equation}
f(\theta,\bar{\theta})=a_{0}+\sum_{n=1}^{k^{\prime}-1}a_{n}\theta^{n}+\sum_{n=1}^{k^{\prime}-1}b_{n}\bar{\theta}^{n}\,.\label{qnormset}\end{equation}
 Consequently a distance and a metric topology can also be defined
on such a subspace.

\section{$k$-fermionic coherent states}

\label{prgcs}

In order to implement the CS quantization for the paragrassmann algebra
$\Sigma_{k}$ we need first to build a family of coherent states.
For that purpose, we follow the scheme described in Section \ref{csq}.
Inspired by \cite{Daoud1998,daoudkibler} we choose as an orthonormal
set in $L_{w}^{2}$ the following monomial functions: \begin{equation}
\phi_{n}(\theta,\bar{\theta})=\frac{\bar{\theta}^{n}}{(\left[n\right]_{q}!)^{\frac{1}{2}}}\,,\quad n=0,1,\dotsc,k^{\prime}-1\,.\label{moncg}\end{equation}

The (nonnormalized) \emph{paragrassmann} or \emph{$k$-fermionic}
coherent states \cite{filisakur,isaev,daoudkibler,rabat1,rabat2}
should be understood as elements of $\Sigma_{k}\otimes\mathbb{C}^{k'}$.
They read as (we omit the tensor product symbol): \begin{equation}
|\theta)=\sum_{n=0}^{k^{\prime}-1}\overline{\phi_{n}(\theta,\bar{\theta})}\,\left|n\right\rangle =\sum_{n=0}^{k^{\prime}-1}\frac{\theta^{n}}{\left(\left[n\right]_{q}!\right)^{\frac{1}{2}}}\,\left|n\right\rangle \,,\label{dkcs13}\end{equation}
 where $\{\left|n\right\rangle \,,\, n=0,1,\dotsc,k^{\prime}-1\}$
is an orthonormal basis of the Hermitian space $\mathbb{C}^{k'}$,
e.g. the canonical basis.

The resolution of the unity $I_{k}$ in $\mathbb{C}^{k'}$ follows
automatically from the orthonormality of the set (\ref{moncg}): \begin{equation}
\int\int d\theta|\theta)\, w(\theta,\bar{\theta})\,(\theta|\, d\bar{\theta}=I_{k}\,.\label{resunGrass13}\end{equation}
 The weight function is given in \eqref{qdefpol13}.

\section{Quantization with $k$-fermionic coherent states}

\label{qtferm}

We now have all the ingredients needed to quantize the set $X=\Sigma_{k}$
of paragrassmann variables along the scheme described in section \ref{csq}.
The quantization of a paragrassmann-valued function $f(\theta,\bar{\theta})$
maps $f$ to the linear operator $A_{f}$ on $\mathbb{C}^{k'}$: \begin{equation}
A_{f}=\int\int d\theta|\theta)\, f(\theta,\bar{\theta})w(\theta,\bar{\theta})\,(\theta|\, d\bar{\theta}\,,\label{quantgrass13}\end{equation}
 with no consideration of ordering, at the moment.

\subsection{Some relations for coherent state quantization of paragrassmann algebras}

Because of the noncommutativity \eqref{ncparagr} between paragrassmann
variables, the quantization proposed in \eqref{quantgrass13} is not
straightforward and one should define an order in which the elements
in the integral should be written. In the following are presented
some possibilities of how to achieve this goal while quantizing a
general function of $\theta$ and $\bar{\theta}$. \begin{equation}
f\left(\theta,\bar{\theta}\right)\mapsto A_{f}=\int d\theta\left|\theta\right):f\left(\theta,\bar{\theta}\right)w\left(\theta,\bar{\theta}\right):\left(\theta\right|d\bar{\theta}\,,\label{qantinormal}\end{equation}
 where the symbol {}``$:\,.\,:$'' stands for antinormal ordering
already adopted in \eqref{inprod}. \begin{eqnarray}
f\left(\theta,\bar{\theta}\right)\overset{L}{\mapsto}A_{f}^{L}=\int d\theta\left|\theta\right)f\left(\theta,\bar{\theta}\right)w\left(\theta,\bar{\theta}\right)\left(\theta\right|d\bar{\theta}\,,\nonumber \\
f\left(\theta,\bar{\theta}\right)\overset{R}{\mapsto}A_{f}^{R}=\int d\theta\left|\theta\right)w\left(\theta,\bar{\theta}\right)f\left(\theta,\bar{\theta}\right)\left(\theta\right|d\bar{\theta}\,.\end{eqnarray}
 Adopting a certain ordering may lead to different quantizations,
for instance to the appearance of an extra $q$-dependent coefficient
inside the summations like we have below for the quantized versions
of $\theta$ and $\bar{\theta}$: \begin{align*}
 & A_{\theta}=A_{\theta}^{L}=\sum_{n=0}^{k^{\prime}-1}\left(\left[n+1\right]_{q}\right)^{1/2}\left|n\right\rangle \left\langle n+1\right|\,,\\
 & A_{\theta}^{R}=\sum_{n=0}^{k^{\prime}-1}\left(\left[n+1\right]_{q}\right)^{1/2}q_{k}^{n+2}\left|n\right\rangle \left\langle n+1\right|\,,\\
 & A_{\bar{\theta}}=A_{\bar{\theta}}^{R}=\sum_{n=0}^{k^{\prime}-1}\left(\left[n+1\right]_{q}\right)^{1/2}\left|n+1\right\rangle \left\langle n\right|\,,\\
 & A_{\bar{\theta}}^{L}=\sum_{n=0}^{k^{\prime}-1}\left(\left[n+1\right]_{q}\right)^{1/2}q_{k}^{n+2}\left|n+1\right\rangle \left\langle n\right|\,.\end{align*}
 Note that, for $f\left(\theta\right)$, $g\left(\theta\right)$,
polynomials in $\theta$, the following algebra homomorphism holds,
\begin{equation}
A_{fg}^{L}=A_{f}^{L}A_{g}^{L}=A_{fg}=A_{f}A_{g}\,.\label{thetahomo}\end{equation}
 Likewise, for $f\left(\bar{\theta}\right)$ and $g\left(\bar{\theta}\right)$
polynomials in $\bar{\theta}$, one has

\[
A_{fg}^{R}=A_{f}^{R}A_{g}^{R}=A_{fg}=A_{f}A_{g}\,.\]

In the following we mainly use the antinormal ordering, already adopted
in Equation \eqref{inprod}. Consequently, we adopt the rule that
$\theta$ is quantized as $A_{\theta}=A_{\theta}^{L}$ and $\bar{\theta}$
as $A_{\bar{\theta}}=A_{\theta}^{R}$.

Now we consider higher order polynomials in $\theta$ and $\bar{\theta}$.
One can easily check that \[
A_{\theta}^{2}=A_{\theta^{2}}\,,\,\, A_{\theta}^{3}=A_{\theta^{3}}\,,\]
 and so on up to order $n=k'-1$. In order to prove this, let us show
that $A_{\theta^{n}}A_{\theta}=A_{\theta^{n+1}}$:\begin{eqnarray*}
A_{\theta^{n}}A_{\theta} & = & \sum_{m,m^{\prime}=0}^{k^{\prime}-1}\left\{ \frac{\left[m+n\right]_{q}!}{\left[m\right]_{q}!}\left[m^{\prime}+1\right]_{q}\right\} ^{1/2}\left|m\right\rangle \left\langle m+n\right.\left|m^{\prime}\right\rangle \left\langle m^{\prime}+1\right|\\
 & = & \sum_{m=0}^{k^{\prime}-1}\left\{ \frac{\left[m+n+1\right]_{q}!}{\left[m\right]_{q}!}\right\} ^{1/2}\left|m\right\rangle \left\langle m+n+1\right|=A_{\theta^{n+1}}\,.\end{eqnarray*}
 This can also be viewed as a direct consequence of the algebra homomorphism
defined in \eqref{thetahomo}; one then ends with the interesting
result $A_{\theta}^{n}=A_{\theta^{n}}$. Similarly, one has $A_{\bar{\theta}}^{n}=A_{\bar{\theta}^{n}}$.
These properties ensure that the nilpotency of the paragrassmann variables
is preserved after quantization.

Calculations regarding the quantization of higher order mixed terms
of $\theta$ and $\bar{\theta}$ are also given in the appendix.

Hence, we recover the $k'\times k'$-matrix realization of the so-called
\emph{$k'$-fermionic algebra} $F_{k'}$ \cite{daoudkibler,rabat1}
(see appendix).

In the simplest case where one quantizes the product $\theta\bar{\theta}$
one gets the interesting formulas:\begin{eqnarray}
A_{\theta\bar{\theta}} & = & \sum_{n=0}^{k^{\prime}-1}\left\vert \left[n+1\right]_{q}\right\vert \;\left|n\right\rangle \left\langle n\right|\nonumber \\
 & = & A_{\theta}^{L}A_{\bar{\theta}}^{R}\;=A_{\theta}A_{\bar{\theta}}\,.\end{eqnarray}
 However \begin{equation}
A_{\bar{\theta}\theta}\neq A_{\bar{\theta}}A_{\theta};\end{equation}
 this is a central point in the discussion that follows.

\subsection{Discussion of the quantized algebra}

\label{qdiscussion}

The quantization of the $q$-commutation relation \eqref{ncparagr}
then is given by: \begin{equation}
A_{\theta}A_{\bar{\theta}}-q\, A_{\bar{\theta}}A_{\theta}=\overline{Q}\label{qcomut}\end{equation}
 where the matrix $\overline{Q}$ is defined by \[
\overline{Q}=\bar{q}^{N}=q^{-N}=\sum_{n=0}^{k^{\prime}-1}\bar{q}^{n}\,\left|n\right\rangle \left\langle n\right|\]
 Equation \eqref{qcomut} is nothing but the defining relation for
the Biedenharn-Macfarlane oscillator \cite{bied,macfarlane}. Another
popular form for equation (\ref{qcomut}) can be obtained by defining
$B_{\theta}=q^{N/2}A_{\theta}$ and $B_{\bar{\theta}}=B_{\bar{\theta}}q^{N/2}$:\[
B_{\theta}B_{\bar{\theta}}-q^{2}B_{\bar{\theta}}B_{\theta}=1\,.\]

In addition, the following rule holds \begin{equation}
A_{\theta}A_{\bar{\theta}}-\bar{q}\, A_{\bar{\theta}}A_{\theta}=Q\,,\label{qcomut2}\end{equation}
 with \[
Q=q^{N}=\sum_{n=0}^{k^{\prime}-1}q^{n}\,\left|n\right\rangle \left\langle n\right|\,.\]

The fact that equations \eqref{qcomut} and \eqref{qcomut2} hold
simultaneously is necessary to ensure that the operators $A_{\theta}$
and $A_{\bar{\theta}}$ are hermitian conjugate to each other. This
actually proved useful while constructing paragrassman coherent states
in \cite{rabat1}. And even though it was an \textit{ad hoc} supposition
then, the present result justifies that choice.

In addition, the fact that $A_{\theta}^{n}=A_{\theta^{n}}$ and $A_{\bar{\theta}}^{n}=A_{\bar{\theta}^{n}}$
ensures the nilpotency of the corresponding quantized operators. This
fact was also used in \cite{rabat2,rabat1} where the constructed
paragrassmann coherent states were associated to nilpotent representations
of Biedenharn-Macfarlane deformed oscillator.

The \textit{q-commutation} relation of the {}``\textit{classical}''
variables $\theta$ and $\bar{\theta}$ : \[
\theta\bar{\theta}-q_{k}\bar{\theta}\theta=0\]
 is replaced by a \textit{non-q-commutativity} relation \eqref{qcomut}
(or \eqref{qcomut2}) between the corresponding quantum operators
$A_{\theta}$ and $A_{\bar{\theta}}$. This is in complete analogy
with the (canonical or coherent states or something else) quantization
of the usual harmonic oscillator where the commutativity of the variables,
\textit{position} and \textit{momentum} is replaced by a noncommutativity
of the corresponding operators in quantum physics.

It is important to note at this stage that these results justify the
choices in works such as \cite{rabat1} where the structure of paragrassmann
algebras was associated with that of the Macfarlane-Biedenharn oscillator.

However, if a different definition of the $q$-numbers \eqref{qdefpolA13}
were adopted one would get the same results, except for those presented
in section \ref{qdiscussion}, where in fact different deformations
of the usual harmonic oscillator are found depending on the definition
adopted for the $q$-numbers \eqref{qdefpolA13}. The common point
between the results is that paragrassmann coherent states are associated
to nilpotent representations of deformations of the harmonic oscillator.

\section{Lower symbols and {}``classical limit''}

\label{lowsymb} The lower symbols for $A_{\theta}$ and $A_{\bar{\theta}}$
are given by the expressions\begin{align*}
 & \left(\theta\right|A_{\theta}\left|\theta\right)=\left(\theta\right.\left|\theta\right)\theta\,,\\
 & \left(\theta\right|A_{\bar{\theta}}\left|\theta\right)=\bar{\theta}\left(\theta\right.\left|\theta\right)\,.\end{align*}
 So, if one can formally define \begin{equation}
\left\langle A_{\theta}\right\rangle \equiv\frac{\left(\theta\right|A_{\theta}\left|\theta\right)}{\left(\theta\right.\left|\theta\right)}=\theta\,,\,\,\left\langle A_{\bar{\theta}}\right\rangle =\frac{\left(\theta\right|A_{\bar{\theta}}\left|\theta\right)}{\left(\theta\right.\left|\theta\right)}=\bar{\theta}\,,\label{qlowsymb}\end{equation}
 then the normalized lower symbols satisfy the classical relation
\[
\left\langle A_{\theta}\right\rangle \left\langle A_{\bar{\theta}}\right\rangle =q_{k}\left\langle A_{\bar{\theta}}\right\rangle \left\langle A_{\theta}\right\rangle \,.\]
 This secures some sort of a returning path, allowing one to recover
the classical algebra from the quantized one using the lower symbols
\eqref{qlowsymb}.

Some useful relations are given in what follows \begin{align*}
 & \theta\left(\theta\right.\left|\theta\right)=\left(\theta\right.\left|q_{k}\theta\right)\theta\,,\,\,\left(\theta\right.\left|\theta\right)\theta=\theta\left(q_{k}\theta\right.\left|\theta\right)\,,\\
 & \bar{\theta}\left(\theta\right.\left|\theta\right)=\left(q_{k}\theta\right.\left|\theta\right)\bar{\theta}\,,\,\,\left(\theta\right.\left|\theta\right)\bar{\theta}=\bar{\theta}\left(\theta\right.\left|q_{k}\theta\right)\,,\\
 & \left(\theta\right.\left|\theta\right)=\left(q_{k}\theta\right.\left|q_{k}\theta\right)=\sum_{n=0}^{k^{\prime}-1}\frac{\left(\bar{\theta}\right)^{n}\left(\theta\right)^{n}}{\left[n\right]_{q}!}=\sum_{n=0}^{k^{\prime}-1}\bar{q}_{k}^{n^{2}}\frac{\left(\theta\right)^{n}\left(\bar{\theta}\right)^{n}}{\left[n\right]_{q}!}\,.\end{align*}

The lower symbol of an arbitrary linear operator $A$ in $\mathbb{C}^{k'}$,
defined by the matrix $(a_{mm'})$ as $A=\sum_{mm'}a_{mm'}\left|m\right\rangle \left\langle m^{\prime}\right|$
is given by the general element of the algebra $\sum_{k}$: \begin{equation}
\left(\theta\right|A\left|\theta\right)=\sum_{n\bar{n}}\frac{a_{\bar{n}n}}{\left(\left[\bar{n}\right]_{q}!\left[n\right]_{q}!\right)^{1/2}}\,\left(\bar{\theta}\right)^{\bar{n}}\left(\theta\right)^{n}=\sum_{n\bar{n}}\frac{\bar{q}_{k}^{n\bar{n}}a_{\bar{n}n}}{\left(\left[\bar{n}\right]_{q}!\left[n\right]_{q}!\right)^{1/2}}\,\left(\theta\right)^{n}\left(\bar{\theta}\right)^{\bar{n}}\,.\label{lowsymbA}\end{equation}

\section{Upper symbols and Moyal product}

\label{upsymb}

The question now is to determine, for an arbitrary linear operator
$A$ in $\mathbb{C}^{k'}$, if there exists $f\in L_{w}^{2}$ such
that $A=A_{f}$, and whether that $f$ is unique or not. It suffices
to show that the coefficients $a_{mn}$ of $A_{f}$, $A_{f}=\sum_{mn}a_{mn}\left|m\right\rangle \left\langle n\right|$,
are uniquely determined by those of $f\left(\theta,\bar{\theta}\right)=\sum_{st}f_{st}\theta^{s}\bar{\theta}^{t}$
and \emph{vice versa}. After performing the Grassmann integrations
in the general expression of the correspondence relation (\ref{quantgrass13}),
one is left with the following relations between the coefficients
of $f$ and $A_{f}$:\begin{equation}
a_{nn^{\prime}}=\sum_{s}M_{nn^{\prime},s}f_{s,n-n^{\prime}+s}\,,\label{eq:a_equals_Mf}\end{equation}
 where\begin{equation}
M_{nn^{\prime},s}=\left(\frac{\left[n+s\right]_{q}!\left[n+s\right]_{q}!}{\left[n\right]_{q}!\left[n^{\prime}\right]_{q}!}\right)^{1/2}\,.\label{eq:Mcoef}\end{equation}
 In order to show that $M$ is invertible, we distinguish two cases,
namely, $n^{\prime}+p$ with non-negative $p\geq0$ and non-positive
$0\geq p$. Since the latter is not independent from the former, we
concentrate on the $p\geqslant0$ case. For $p\geqslant0$, we write
(\ref{eq:a_equals_Mf}) as\[
a_{n}^{\left(p\right)}=\sum_{s=0}^{k^{\prime}-1-n}M_{n,s}^{\left(p\right)}f_{s}^{\left(p\right)}\,,\]
 with the obvious identifications $a_{n}^{\left(p\right)}\equiv a_{n,n-p}$,
$M_{n,s}^{\left(p\right)}=M_{n,n-p,s}$ and $f_{s}^{\left(p\right)}=f_{s,p+s}$.
The upper summation limit arises from the restriction in $M_{n,s}^{\left(p\right)}$
that $n+s<k'$ and $n\geq p$. The linear system thus obtained can
be better understood in terms of the matrices\[
\left(\begin{array}{c}
a_{k^{\prime}-1}^{\left(p\right)}\\
a_{k^{\prime}-2}^{\left(p\right)}\\
\vdots\\
a_{p}^{\left(p\right)}\end{array}\right)=\left(\begin{array}{cccc}
M_{k^{\prime}-1,0}^{\left(p\right)} & 0 & \cdots & 0\\
M_{k^{\prime}-2,0}^{\left(p\right)} & M_{k^{\prime}-2,1}^{\left(p\right)} & \ddots & 0\\
\vdots & \vdots & \ddots & 0\\
M_{p,0}^{\left(p\right)} & M_{p,1}^{\left(p\right)} & \cdots & M_{p,k^{\prime}-1-p}^{\left(p\right)}\end{array}\right)\left(\begin{array}{c}
f_{0}^{\left(p\right)}\\
f_{1}^{\left(p\right)}\\
\vdots\\
f_{k^{\prime}-1-p}^{\left(p\right)}\end{array}\right)\,.\]
 Note that for each integer $s$ in $k^{\prime}-1-p\geq s\geq0$,
the coefficient $M_{k^{\prime}-1-s,s}^{\left(p\right)}$of $f_{s}^{\left(p\right)}$
in $a_{k^{\prime}-1-s}^{\left(p\right)}$ is nonzero. This means that
the above triangular matrix can be inverted, since its determinant
is $\Pi_{s=0}^{k^{\prime}-1-p}M_{k^{\prime}-1-s,s}^{\left(p\right)}\neq0$.
The case with $p<0$ leads to an analogous system with identical non-singular
triangular matrix consisting of the coefficients (\ref{eq:Mcoef}).

Therefore, we have established that the correspondence $f\mapsto A_{f}$
is invertible. For instance, for $k=4$, the inverse image of any
order $2$ matrix $a_{mn}$ is\begin{equation}
f\left(\theta,\bar{\theta}\right)=a_{11}+a_{01}\theta+a_{10}\bar{\theta}+\left(a_{00}-a_{11}\right)\theta\bar{\theta}\,.\label{eq:k2symbol}\end{equation}

As a result of the unique correspondence $f\mapsto A_{f}$, the algebra
of matrices on $\mathbb{C}^{k'}$ is isomorphic to the algebra of
functions $f\left(\theta,\bar{\theta}\right)$ with Moyal multiplication:\[
A_{f}\, A_{g}=A_{f\star g}\,.\]
 As an example, let us consider the $k=4$ case and an application
to the complex quaternion algebra . Given $f\mapsto A_{f}$ and $g\mapsto A_{g}$,
the Moyal product of the symbols $f$ and $g$ of the operators $A_{f}$
and $A_{g}$ is\begin{align}
f\star g\left(\theta,\bar{\theta}\right) & =f_{01}g_{10}+f_{00}g_{00}+\left(\left(f_{00}+f_{11}\right)g_{10}+f_{10}g_{00}\right)\theta+\left(f_{01}\left(g_{00}+g_{11}\right)+f_{00}g_{01}\right)\bar{\theta}\nonumber \\
 & +\left(f_{00}g_{11}+f_{11}\left(g_{00}+g_{11}\right)+f_{10}g_{01}-f_{01}g_{10}\right)\theta\bar{\theta}\,.\label{eq:k2moyalprod}\end{align}
 One can easily see that $\theta\star\theta=\bar{\theta}\star\bar{\theta}=0$.
Now consider the (Pauli) matrix representation for the complex quaternions,
\[
Z=z_{0}+z_{1}\left(i\sigma^{1}\right)+z_{2}\left(-i\sigma^{2}\right)+z_{3}\left(i\sigma^{3}\right)\,,\,\, z_{i}\in\mathbb{C}\,.\]
 One can use formulae (\ref{eq:k2symbol},\ref{eq:k2moyalprod}) to
compute the lower symbol of the product $ZW$, where $Z$ and $W$
are as above. As expected, the Moyal product reproduces the quaternion
multiplication law in symbol space, for the complex coefficients $x_{i}$
defining $z\star w$ are\[
x_{0}=z_{0}w_{0}-\mathbf{z}\cdot\mathbf{w}\,,\,\,\mathbf{x}=z_{0}\mathbf{w}+w_{0}\mathbf{z}+\mathbf{z}\times\mathbf{w}\,.\]
 Equivalently, one can compute the symbols $i\sigma^{1}\mapsto I$,
$-i\sigma^{2}\mapsto J$ and $i\sigma^{3}\mapsto K$ and easily see
that \begin{align*}
 & I\star I=J\star J=K\star K=-1\,,\\
 & I\star J=-J\star I=K\end{align*}
 \,. Thus, the general symbol according to (\ref{eq:k2symbol}) is\begin{align*}
z\left(\theta,\bar{\theta}\right) & =z_{0}-iz_{3}+\left(iz_{1}-z_{2}\right)\theta+\left(iz_{1}+z_{2}\right)\bar{\theta}+2iz_{3}\theta\bar{\theta}\\
 & =z_{0}+z_{1}I+z_{2}J+z_{3}K\,.\end{align*}
 from which is evident that $z\star w$ is quaternion multiplication.

\section{Paragrassmannian Fock-Bargmann representation of operators}

To any vector $\left|\psi\right\rangle $ in $\mathbb{C}^{k'}$ let
us associate the function $\psi(\theta)$ in $L_{w}^{2}$ defined
by

\begin{equation}
\mathcal{W}:\left|\psi\right\rangle \mapsto\psi(\theta)\overset{\mathrm{def}}{=}\left(\theta\right|\psi\rangle=\sum_{n=0}^{k^{\prime}-1}\frac{\theta^{n}}{(\left[n\right]_{q}!)^{\frac{1}{2}}}\,\psi_{n}\,,\quad\psi_{n}\equiv\langle n\left|\psi\right\rangle \,,\label{fbrep}\end{equation}
 This yields an isometry between $\mathbb{C}^{k'}$ and the Hilbert
subspace of $L_{w}^{2}$ spanned by the first $k$ powers of $\theta$.
Under such an isometry, the operators $A_{\theta}$ and $A_{\bar{\theta}}$
are denoted respectively by: \begin{equation}
\partial_{\theta}\overset{\mathrm{def}}{=}\mathcal{W}A_{\theta}\mathcal{W}^{\dagger}\,\quad\mathfrak{m}_{\theta}\overset{\mathrm{def}}{=}\mathcal{W}A_{\bar{\theta}}\mathcal{W}^{\dagger}\,.\label{fbdelta}\end{equation}
 They realize themselves, respectively, as paragrassmannian derivative
and multiplication as: \begin{equation}
\partial_{\theta}\,\theta^{n}=[n]_{q}\,\theta^{n-1}\,,\quad\mathfrak{m}_{\theta}\, f(\theta)=\theta f(\theta)\,.\label{realfbdelta}\end{equation}
 Hence, this {}``Fock-Bargmann'' representation of operators allows
one to recover in a quite natural way the algebra of type $\Pi_{k}$
with two nilpotent generators $\theta$ and $\partial$ as they are
described in \cite{isaev}.

\section{Generalization to $d$-dimensional paragrassmann variable algebra}

\label{dvarprg}

All the results obtained so far can be generalized to the $d$-dimensional
case for which the observation set, $X$, is given by a $d$-dimensional
paragrassmann algebra.

A $d$-dimensional paragrassmann algebra, \cite{majrod}, is generated
by $d$ paragrassmann variables $\theta_{i}$ and their respective
conjugates $\bar{\theta}_{i}$ where $i=1,2,...,d$. The variables
do not commute with each other. One has instead the following $q$-commutation
relations: \begin{equation}
\begin{array}{cclc}
\theta_{i}\theta_{j} & = & q_{k}\;\theta_{j}\theta_{i}\;,\;\;\;\;\\
\bar{\theta}_{i}\bar{\theta}_{j}\,, & = & q_{k}\bar{\theta}_{j}\bar{\theta}_{i}\,, & i,j=1,2,\ldots,d\;,\;\; i<j\,.\\
\theta_{i}\bar{\theta}_{j} & = & \bar{q}_{k}\bar{\theta}_{j}\theta_{i}\,.\end{array}\label{qdtheta}\end{equation}
 Here $q_{k}$ is a $k$th root of unity for odd $k$ and a $\frac{k}{2}$th
root of unity for even $k$, as before. All these variables are nilpotent:
$\theta_{i}^{k^{\prime}}=\bar{\theta}_{i}^{k^{\prime}}=0$. Relations
(\ref{qdtheta}) can be written in a compact form as follows \begin{equation}
\alpha_{i}\beta_{j}=q_{k}^{ab}\beta_{j}\alpha_{i}\;\;\;\; i<j\,,\quad a,b\in\{-1,1\}.\label{majidgrassmann}\end{equation}
 A measure on the observation set, analogous to the one in \eqref{dkmeas13},
is given by: \begin{equation}
\mu(d\theta d\bar{\theta})=d\theta_{d}...d\theta_{1}\, w(\theta,\bar{\theta})\, d\bar{\theta_{d}}...d\bar{\theta_{1}}\,.\label{dthetameas}\end{equation}

We use here the short-handed notation $w(\theta,\bar{\theta})$ for
the weight, which is in fact a polynomial function of all the variables:
\begin{eqnarray}
w(\theta,\bar{\theta})=\sum_{n_{1},n_{2},...,n_{d}=0}^{k^{\prime}-1} &  & \left[n_{1}\right]_{q}!\,\left[n_{2}\right]_{q}!\,...\,\left[n_{d}\right]_{q}!\\
 &  & \theta_{1}^{k^{\prime}-1-n_{1}}\,\theta_{2}^{k^{\prime}-1-n_{2}}\,...\,\theta_{d}^{k^{\prime}-1-n_{d}}\,\bar{\theta}_{1}^{k^{\prime}-1-n_{1}}\,\bar{\theta}_{2}^{k^{\prime}-1-n_{2}}\,...\,\bar{\theta}_{d}^{k^{\prime}-1-n_{d}}\,.\nonumber \end{eqnarray}
 The integration is also carried out in the sense of Berezin-Majid-Rodríguez-Plaza
integrals \textit{i.e.} Eqs. \eqref{bermajint13} and \eqref{bermajintA13}
hold for each of the variables $\theta_{i}$ and $\bar{\theta}_{i}$.

We define $d$-paragrassmann coherent states as tensor products of
$d$ single mode paragrassmann coherent states given in Eq. \eqref{dkcs13}:

\begin{eqnarray}
\vert\theta) & = & \vert\theta_{1}\,\theta_{2}\,...\,\theta_{d})=\vert\theta_{1})\otimes\vert\theta_{2})\otimes\,...\,\otimes\vert\theta_{d})\\
 & = & \sum_{n_{1},n_{2},...,n_{d}=0}^{k^{\prime}-1}\frac{\theta_{1}^{n_{1}}\,\theta_{2}^{n_{2}}\,...\,\theta_{d}^{n_{d}}}{\left(\left[n_{1}\right]_{q}!\,\left[n_{2}\right]_{q}!\,...\,\left[n_{d}\right]_{q}!\right)^{1/2}}\;\vert n_{1},\, n_{2},\,...\, n_{d}\rangle\,.\end{eqnarray}

Together with the measure in \eqref{dthetameas}, these states provide
us with a resolution of the unity in the Hilbert space $(\mathbb{C}^{k'})^{\otimes d}$:
\begin{equation}
\int d\theta_{d}...d\theta_{1}\,::\,\vert\theta)\, w(\theta,\bar{\theta})(\theta\vert\,::\, d\bar{\theta_{d}}...d\bar{\theta_{1}}=I\,.\label{dresolution}\end{equation}
 Here also, because of the noncommutativity of the paragrassmann variables,
an ordering should be adopted. In Eq. \eqref{dresolution} the ::
:: stands for an ordering in which all the $\theta$'s are to the
left of the $\bar{\theta}$'s, and the $\theta$'s (as well as the
$\bar{\theta}$'s) are ordered according to their indices in \underbar{increasing}
order.

Let us use the notation $f(\theta,\bar{\theta})$ for a generic polynomial
function of all the variables $\theta_{i}$ and $\bar{\theta}_{i}$.
Then following the quantization scheme described in Section \ref{csq},
such a a generic function can be quantized and mapped to an operator
$A_{f}$ acting on $(\mathbb{C}^{k'})^{\otimes d}$. The corresponding
operator is given by \begin{equation}
A_{f}=\int d\theta::\, f\;\vert\theta)\, w(\theta,\bar{\theta})(\theta\vert\,::\, d\bar{\theta}\,,\end{equation}
 where the following short-handed notations are adopted: \[
d\theta=d\theta_{d}...d\theta_{1}\;\;\text{and}\;\; d\bar{\theta}=d\bar{\theta_{d}}...d\bar{\theta_{1}}\,.\]

For the simplest functions we get the following results: \begin{eqnarray}
A_{\theta_{i}} & = & \int d\theta::\,\theta_{i}\;\vert\theta)\, w(\theta,\bar{\theta})(\theta\vert\,::\, d\bar{\theta}\nonumber \\
 & = & \sum_{n_{1},n_{2},...,n_{d}=0}^{k^{\prime}-1}\left(\left[n_{i}+1\right]_{q}\right)^{1/2}\vert n_{1},\, n_{2},\,...\, n_{d}\rangle\langle n_{1},\, n_{2},\,...,n_{i}+1,\,...,\, n_{d}\vert\,,\end{eqnarray}

\begin{equation}
A_{\bar{\theta}_{j}}=\sum_{n_{1},n_{2},...,n_{d}=0}^{k^{\prime}-1}\left(\left[n_{j}+1\right]_{q}\right)^{1/2}\vert n_{1},\, n_{2},\,...,n_{j}+1,\,...,\, n_{d}\rangle\langle n_{1},\, n_{2},\,...\, n_{d}\vert\,.\end{equation}

We recognize in $A_{\theta_{i}}$ (respectively, $A_{\bar{\theta}_{i}}$)
the lowering or annihilation operator (respectively, raising or creation
operator) in the $i^{th}$ mode. The product of two such operators
is given by: \begin{align}
 & A_{\theta_{i}}A_{\bar{\theta}_{j}}=\nonumber \\
 & \sum_{n_{1},n_{2},...,n_{d}=0}^{k^{\prime}-1}\left(\left[n_{i}+1\right]_{q}\left[n_{j}+1\right]_{q}\right)^{1/2}\vert n_{1},\, n_{2},\,...,n_{j}+1,\,...,\, n_{d}\rangle\langle n_{1},\, n_{2},\,...,n_{i}+1,\,...,\, n_{d}\vert\,.\end{align}
 and \begin{align}
 & A_{\bar{\theta}_{j}}A_{\theta_{i}}=\nonumber \\
 & \sum_{n_{1},n_{2},...,n_{d}=0}^{k^{\prime}-1}\left(\left[n_{i}+1\right]_{q}\left[n_{j}+1\right]_{q}\right)^{1/2}\vert n_{1},\, n_{2},\,...,n_{j}+1,\,...,\, n_{d}\rangle\langle n_{1},\, n_{2},\,...,n_{i}+1,\,...,\, n_{d}\vert\,.\end{align}

For $i=j$ we derive a result similar to the one-variable case \eqref{qcomut}:
\begin{equation}
A_{\theta_{i}}A_{\bar{\theta}_{i}}-q\, A_{\bar{\theta}_{i}}A_{\theta_{i}}=\overline{Q}_{i}\,,\label{idqcomut}\end{equation}
 where \begin{equation}
\overline{Q}_{i}=\bar{q}^{N_{i}}=q^{-N_{i}}=\sum_{n_{1},n_{2},...,n_{d}=0}^{k^{\prime}-1}\bar{q}^{n_{i}}\,\vert n_{1},\, n_{2},\,...,\, n_{d}\rangle\langle n_{1},\, n_{2},\,...\, n_{d}\vert\,.\end{equation}

Eq. \eqref{idqcomut} is the quantized version of the \textit{$q$-commutativity}
expressed in the last of the equations in \eqref{qdtheta}. The quantization
process breaks this \textit{$q$-commutativity} and replaces it by
the non\textit{-$q$-commutativity} \eqref{idqcomut}.

For $i<j$ the corresponding operators commute \begin{equation}
A_{\theta_{i}}A_{\bar{\theta}_{j}}-A_{\bar{\theta}_{j}}A_{\theta_{i}}=0\,.\label{ijcommut}\end{equation}
 In this case, the quantization process broke the \textit{$q$-commutativity}
and replaced it by ordinary commutative relations.

Now let us quantize higher order terms of $\theta$'s: \begin{equation}
A_{\theta_{i}\theta_{j}}=\sum_{n_{1},n_{2},...,n_{d}=0}^{k^{\prime}-1}\,\left(\left[n_{i}+1\right]_{q}\left[n_{j}+1\right]_{q}\right)^{1/2}\vert n_{1},\, n_{2},\,...\, n_{d}\rangle\langle n_{1},\,...,n_{i}+1,\,...,n_{j}+1\,,...\, n_{d}\vert\,;\end{equation}
 this holds for $i\neq j$ and we have: \begin{equation}
A_{\theta_{i}\theta_{j}}=A_{\theta_{i}}A_{\theta_{j}}=A_{\theta_{j}\theta_{i}}=A_{\theta_{j}}A_{\theta_{i}}\,.\label{ijthetacomut}\end{equation}
 This result is similar to Eq. \eqref{ijcommut}: the \textit{$q$-commutativity}
in the first equation in \eqref{qdtheta} is broken and replaced by
\textit{commutativity}. Eq. \eqref{ijthetacomut} endows another feature:
to, apparently, two different classical functions, $\theta_{i}\theta_{j}$
and $\theta_{j}\theta_{i}$ there corresponds the same quantum operator
$A_{\theta_{i}\theta_{j}}=A_{\theta_{j}\theta_{i}}$. This feature
is a generic consequence of the adopted antinormal and index ordering
in \eqref{dresolution}.

For $i=j$ we have \begin{equation}
A_{\theta_{i}\theta_{i}}=\sum_{n_{1},n_{2},...,n_{d}=0}^{k^{\prime}-1}\,\left(\frac{\left[n_{i}+2\right]_{q}!}{\left[n_{i}\right]_{q}!}\right)^{1/2}\vert n_{1},\, n_{2},\,...\, n_{d}\rangle\langle n_{1},...,n_{i}+2,...\, n_{d}\vert\,,\end{equation}
 and we have the same homomorphism for higher order terms of $\theta$
and $\bar{\theta}$: \begin{equation}
A_{\theta_{i}}^{n}=A_{\theta_{i}^{n}}\,,\,\, A_{\bar{\theta}_{i}}^{n}=A_{\bar{\theta}_{i}^{n}}\,.\end{equation}
 So we have the same nilpotency conditions for the quantized operators
associated with paragrassmann variables: \[
A_{\theta_{i}}^{k^{\prime}}=A_{\bar{\theta}_{i}}^{k^{\prime}}=0\,.\]
 These operators could be useful for the description of \textit{parafermions
\cite{ruspi}} or $k$-\emph{fermions} \cite{daoudkibler}, which
are hypothetical particles obeying a sort of generalized Pauli's exclusion
principle: the maximum number of parafermions allowed to occupy the
same quantum state is $k-1$ (1 for fermions).

If one would follow this reasoning and use the $d$-dimensional paragrassmann
algebra together with the associated quantized algebra to describe
such particles, the operators $A_{\bar{\theta}_{i}}$ could be interpreted
as the creation operator in the $i^{th}$ mode, while $A_{\theta_{i}}$
would annihilate a particle from the same mode. What we learn from
the commutation relations between the quantized operators is that
particles in different modes do not interact with each other; this
is the essence of the following commutation relations: \begin{eqnarray}
[A_{\theta_{i}}\;,\; A_{\theta_{j}}]=0\,,\quad &  & \quad[A_{\bar{\theta}_{i}}\;,\; A_{\bar{\theta}_{j}}]=0\,,\nonumber \\
\left[A_{\theta_{i}}\;,\; A_{\bar{\theta}_{j}}\right] & = & 0\,,\end{eqnarray}
 while in the same mode the operators obey the non-$q$-\textit{commutativity}
relation in \eqref{idqcomut}.

\section{Conclusion}

\label{conc}In this work we have implemented a coherent-state quantization
of the paragrassmann algebra $\Sigma_{k}$ viewed as a {}``classical''
phase space in a wide sense. The followed procedure has yielded a
unique correspondence between $\Sigma_{k}$ and the $k'\times k'$
matrix realization of the so-called \emph{$k$-fermionic algebra}
$F_{k}$ \cite{daoudkibler,rabat1,daoud1997}. In particular, the
$q$-commutation relations between paragrassmann generators are mapped
to the Biedenharn-Macfarlane commutation relations for the $q$-oscillator
\cite{bied,macfarlane}. Moreover, the properties of nilpotency and
hermiticity of the matrix operators $A_{\theta}$ and $A_{\bar{\theta}}$,
which are required in view of the construction of paragrassman coherent
states in \cite{rabat1,rabat2}, arise naturally in our construction.

We also note that, as a result of the uniqueness of the correspondence
$f\left(\theta,\bar{\theta}\right)\mapsto A_{f}$, one can reexpress
any finite-dimensional quantum algebra in terms of the Moyal algebra
of the corresponding symbols.

We believe that our approach to quantizing paragrassmann algebras
is suitable for the investigation of classical systems of particles
with paragrassmann degrees of freedom and their quantum analogs, in
the spirit of the Berezin-Marinov particle models \cite{Marinov1977}.
We also think that the displayed one-to-one correspondence between
finite-dimensional matrix algebra and a more elaborate algebraic structure
sheds a new light on the question of the equivocal nature of what
we call quantization. Finally, we think that it would be of great
interest to explore such {}``classical''-quantum correspondence
within the quantum information context by extending our formalism
to tensor products of quantum states and studying their respective
{}``classical'' paragrassmanian counterparts.

\section*{Acknowledgments}

The authors thank M. Kibler for very helpful comments and suggestions
on the content of this paper.

\section{Appendix}

\label{ap1}

\textbf{Commutation relations between first-order elements:}

\begin{align*}
 & A_{\theta}^{L}A_{\bar{\theta}}^{R}=\sum_{n=0}^{k^{\prime}-1}\left(\left[n+1\right]_{q}\left[n+1\right]_{q}\right)^{1/2}\left|n\right\rangle \left\langle n\right|\\
 & A_{\theta}^{L}A_{\bar{\theta}}^{L}=A_{\theta}^{R}A_{\bar{\theta}}^{R}=\sum_{n=0}^{k^{\prime}-1}\left(\left[n+1\right]_{q}\left[n+1\right]_{q}\right)^{1/2}q_{k}^{n+2}\left|n\right\rangle \left\langle n\right|\\
 & A_{\theta}^{R}A_{\bar{\theta}}^{L}=\sum_{n=0}^{k^{\prime}-1}\left(\left[n+1\right]_{q}\left[n+1\right]_{q}\right)^{1/2}q_{k}^{2n+4}\left|n\right\rangle \left\langle n\right|\end{align*}
 \begin{align*}
 & A_{\bar{\theta}}^{R}A_{\theta}^{L}=\sum_{n=0}^{k^{\prime}-1}\left(\left[n\right]_{q}\left[n\right]_{q}\right)^{1/2}\left|n\right\rangle \left\langle n\right|\,,\,\, A_{\bar{\theta}}^{R}A_{\theta}^{R}=A_{\bar{\theta}}^{L}A_{\theta}^{L}=\sum_{n=0}^{k^{\prime}-1}\left(\left[n\right]_{q}\left[n\right]_{q}\right)^{1/2}q_{k}^{n+1}\left|n\right\rangle \left\langle n\right|\\
 & A_{\bar{\theta}}^{L}A_{\theta}^{R}=\sum_{n=0}^{k^{\prime}-1}\left(\left[n\right]_{q}\left[n\right]_{q}\right)^{1/2}q_{k}^{2n+2}\left|n\right\rangle \left\langle n\right|\end{align*}

\textbf{Higher order mixed terms}

Now let us consider mixed terms such as $\theta^{n}\bar{\theta}^{m}$
with $n>m$.\begin{align}
A_{\theta^{n}\bar{\theta}^{m}}=A_{\bar{\theta}^{m}\theta^{n}} & =\int d\theta\left|\theta\right)\theta^{n}w\left(\theta,\bar{\theta}\right)\bar{\theta}^{m}\left(\theta\right|d\bar{\theta}=\sum_{l=0}^{k^{\prime}-1}\left\{ \frac{\left[l+n\right]_{q}!}{\left[l\right]_{q}!}\frac{\left[l+n\right]_{q}!}{\left[l+n-m\right]_{q}!}\right\} ^{1/2}\left|l\right\rangle \left\langle l+n-m\right|\nonumber \\
 & =A_{\theta^{n}}A_{\bar{\theta}^{m}}\,.\end{align}
 It is not true, however, that $A_{\theta^{n}}A_{\bar{\theta}^{m}}=A_{\bar{\theta}^{m}}A_{\theta^{n}}$:\[
A_{\bar{\theta}^{m}}A_{\theta^{n}}=\sum_{l=0}^{k^{\prime}-1}\left\{ \frac{\left[l+n\right]_{q}!}{\left[l\right]_{q}!}\frac{\left[l+m\right]_{q}!}{\left[l\right]_{q}!}\right\} ^{1/2}\left|l+m\right\rangle \left\langle l+n\right|\]
 Note that this result is also valid for $n=m$. In the case $m>n$,
one has \begin{equation}
A_{\theta^{n}\bar{\theta}^{m}}=A_{\bar{\theta}^{m}\theta^{n}}=\sum_{l=0}^{k^{\prime}-1}\left\{ \frac{\left[l+m\right]_{q}!}{\left[l+m-n\right]_{q}!}\frac{\left[l+m\right]_{q}!}{\left[l\right]_{q}!}\right\} ^{1/2}\left|l+m-n\right\rangle \left\langle l\right|=A_{\theta^{n}}A_{\bar{\theta}^{m}}\,.\label{quantmon2}\end{equation}
 One can write \[
A_{\theta}A_{\bar{\theta}^{m}}=A_{\theta}A_{\bar{\theta}}A_{\bar{\theta}^{m-1}}=\left[A_{\theta},A_{\bar{\theta}}\right]A_{\bar{\theta}^{m-1}}+A_{\bar{\theta}}A_{\theta}A_{\bar{\theta}^{m-1}}\]
 and successively, so that\[
\left[A_{\theta},A_{\bar{\theta}^{m-1}}\right]=\sum_{r=0}^{m-1}A_{\bar{\theta}^{r}}\left[A_{\theta},A_{\bar{\theta}}\right]A_{\bar{\theta}^{m-1-r}}\]
 which corresponds to total symmetrization of $A_{\bar{\theta}^{m}}$
around the commutator. By iterating the expression \[
A_{\theta^{n}}A_{\bar{\theta}^{m}}=A_{\theta}A_{\theta^{n-1}}A_{\bar{\theta}^{m}}=A_{\theta}\left[A_{\theta^{n-1}},A_{\bar{\theta}^{m}}\right]+\left[A_{\theta},A_{\bar{\theta}^{m}}\right]A_{\theta^{n-1}}+A_{\bar{\theta}^{m}}A_{\theta^{n}}\]
 one obtains \[
\left[A_{\theta^{n}},A_{\bar{\theta}^{m}}\right]=\sum_{r=0}^{n-1}A_{\theta^{r}}\left[A_{\theta},A_{\bar{\theta}^{m}}\right]A_{\theta^{n-1-r}}=\sum_{s=0}^{n-1}\sum_{r=0}^{m-1}A_{\theta^{s}}A_{\bar{\theta}^{r}}\left[A_{\theta},A_{\bar{\theta}}\right]A_{\bar{\theta}^{m-1-r}}A_{\theta^{n-1-s}}\]

\paragraph{The $k$-fermionic algebra $F_{k}$}

Quoting section 2.1 from \cite{daoudkibler}, we define here the $k$-fermionic
algebra $F_{k}$. {}``The algebra $F_{k}$ is spanned by five operators
$f_{-}$, $f_{+}$, $f_{+}^{+}$, $f_{-}^{+}$ and $N$ through the
following relations classified in three types.

\noindent (i) The $[f_{-},f_{+},N]$-type: \[
f_{-}f_{+}-qf_{+}f_{-}=1\]
 \[
Nf_{-}-f_{-}N=-f_{-},\quad Nf_{+}-f_{+}N=+f_{+}\]
 \[
\left(f_{-}\right)^{k}=\left(f_{+}\right)^{k}=0\]
 (ii) The $[f_{+}^{+},f_{-}^{+},N]$-type: \[
f_{+}^{+}f_{-}^{+}-\bar{q}f_{-}^{+}f_{+}^{+}=1\]
 \[
Nf_{+}^{+}-f_{+}^{+}N=-f_{+}^{+},\quad Nf_{-}^{+}-f_{-}^{+}N=+f_{-}^{+}\]
 \[
\left(f_{+}^{+}\right)^{k}=\left(f_{-}^{+}\right)^{k}=0\]
 (iii) The $[f_{-},f_{+},f_{+}^{+},f_{-}^{+}]$-type: \[
f_{-}f_{+}^{+}-q^{-\frac{1}{2}}f_{+}^{+}f_{-}=0,\quad f_{+}f_{-}^{+}-q^{+\frac{1}{2}}f_{-}^{+}f_{+}=0\]
 where the number \[
q:={\rm exp}\left(\frac{2\pi{\rm i}}{k}\right),\quad k\in{\bf N}\setminus\{0,1\}\]
 is a root of unity and $\bar{q}$ stands for the complex conjugate
of $q$. The couple $(f_{-},f_{+}^{+})$ of annihilation operators
is connected to the couple $(f_{+},f_{-}^{+})$ of creation operators
via the Hermitean conjugation relations \[
f_{+}^{+}=\left(f_{+}\right)^{\dagger},\quad f_{-}^{+}=\left(f_{-}\right)^{\dagger}\]
 and $N$ is an Hermitean operator. It is clear that the case $k=2$
corresponds to fermions and the case $k\to\infty$ to bosons. In the
two latter cases, we can take $f_{-}\equiv f_{+}^{+}$ and $f_{+}\equiv f_{-}^{+}$.
In the other cases, the consideration of the two couples $(f_{-},f_{+}^{+})$
and $(f_{+},f_{-}^{+})$ is absolutely necessary. In the case where
$k$ is arbitrary, we shall speak of $k$-fermions.''

\end{document}